\documentclass[twocolumn,aps,superscriptaddress,pre]{revtex4}

\usepackage{amsmath,amssymb,graphicx}
\usepackage{algorithmic}
\usepackage{enumerate}
\usepackage{times}
\usepackage{color}
\usepackage{soul}
\definecolor{yblue}{rgb}{0.06, 0.3, 0.57}
\usepackage[pdftex]{hyperref}
\hypersetup{colorlinks=true,linkcolor=blue,citecolor=blue,urlcolor=blue}

\begin{document}

\title{Melting of a two-dimensional monodisperse cluster crystal to a cluster liquid}
\date{\today}

\author{Wenlong Wang}
\email{wenlongcmp@gmail.com}
\affiliation{Department of Physics, Royal Institute of Technology, Stockholm, SE-106 91, Sweden}

\author{Rogelio D\'iaz-M\'endez}
\affiliation{Department of Physics, Royal Institute of Technology, Stockholm, SE-106 91, Sweden}

\author{Mats Wallin}
\affiliation{Department of Physics, Royal Institute of Technology, Stockholm, SE-106 91, Sweden}

\author{Jack Lidmar}
\affiliation{Department of Physics, Royal Institute of Technology, Stockholm, SE-106 91, Sweden}

\author{Egor Babaev}
\affiliation{Department of Physics, Royal Institute of Technology, Stockholm, SE-106 91, Sweden}

\begin{abstract}
Monodisperse ensembles of particles that have cluster crystalline phases at low temperatures can model a number of physical systems, such as vortices in type-1.5 superconductors, colloidal suspensions and cold atoms.  In this work we study a two-dimensional cluster-forming particle system interacting via an ultrasoft potential. We present a simple mean-field characterization of the cluster-crystal ground state, corroborating with Monte Carlo simulations for a wide range of densities.  The efficiency of several Monte Carlo algorithms are compared and the challenges of thermal equilibrium sampling are identified.  We demonstrate that the liquid to cluster-crystal phase transition is of first order and occurs in a single step, and the liquid phase is a cluster liquid.  
\end{abstract}

\maketitle

\section{Introduction}
Melting of two-dimensional crystals is a fundamentally interesting problem that attracted interest for decades.  Within the celebrated Kosterlitz-Thouless-Halperin-Nelson-Young (KTHNY) theory~\cite{Halperin:Melting,nelson79}, the melting is a two-stage process occurring with an intermediate hexatic phase of quasi-long range orientational order separating the crystal from the fluid.  In this scenario, the two phase transitions (crystal-to-hexatic and hexatic-to-fluid) are predicted to be of second order, driven by the unbinding of different types of topological defects (dislocations and disclinations)~\cite{nelson79,kosterlitz16}.  However, after many decades of experimental and numerical research, evidence was collected that several 2D systems fall outside KTHNY theory, melting in a single first order transition, or in a two-step process by one first order and one continuous transitions  \cite{armstrong89,bernard11,kapfer15}.  While the characterization of melting transitions of 2D crystals is a long-standing problem that has received great attention, almost all the effort has been focused on simple potentials forming simple crystals, with only a few numerical studies available on the generalized exponential model~\cite{prestipino14,montes15}.

Monodisperse systems, in which particles are all of the same type, can nonetheless exhibit very complex
hierarchical structure formations such as lattices of clusters and exotic phases such as glasses
\cite{MPstripe,Glaser:SS,RRphases,Ikeda:Glass,Rogelio:Glass,zu17,CCSH}.  The wide array of behaviors including rich self-organizing patterns, dynamics and the thermodynamic equilibrium phases in such systems are far from  understood~\cite{ca:hertz,ca:patterns,kobi:quasi,Rogelio:Cluster}.  Particles with cluster-forming interactions emerge in various contexts ranging from soft matter, complex molecules, to cold atoms and vortices in superconductors~\cite{cinti14,meng16}. The problem is also relevant for the physics of multi-component superconductors where a rich variety of vortex cluster solutions have been found \cite{meng2014honeycomb,varney2013hierarchical,Rogelio:Glass}.

For non-cluster-forming systems, the ordered phase consists in a normal crystal with a single particle located at each lattice site.  On the other hand, for cluster-forming potentials the particles arrange in a cluster crystal, in which each lattice site is occupied by a cluster of more than one particle.  Under the above conditions, there is a simple criterion by Likos \textit{et al.} relating the cluster-forming ability of the interactions to the sign of its Fourier transform~\cite{Likos:Criteria,likos07}.  Namely, if the Fourier transform of the interaction potential is positive definite, the system is non-cluster-forming, and  cluster-forming otherwise.  

Since the ordered states of cluster-forming particles in principle allow more types of defects, this raises the question of the nature of melting transition of cluster-crystals in 2D, and to what extent the KTHNY theory applies. In this work, we focus on a cluster-forming ultrasoft potential that has attracted considerable attention as a model for colloidal suspensions and ultracold atoms \cite{cinti14,Rogelio:Glass}.  Recently a similar kind of cluster-forming interaction potential was found to arise between the vortices in type-1.5 superconductors. In the bulk, intervortex forces are long-range attractive and short-range repulsive; see e.g. 
Refs.~\cite{Babaev.Speight:05,Babaev.Carlstrom.ea:10,Silaev.Babaev:11,babaev2017type}. For thin films an additional repulsive interaction arises due to magnetic stray fields that gives raise to vortex cluster crystals \cite{varney2013hierarchical,Rogelio:Glass}. We present a simple mean-field analysis of the ordered cluster-crystalline ground state.  In particular, our analysis captures a number of interesting properties such as the existence of an onset density for the cluster formation, the density dependence of the lattice constant, and also that the triangular lattice is more stable than the square lattice at all densities.  We corroborate our results with a number of Monte Carlo simulations, finding a qualitative and also reasonable quantitative agreement.

In order to ensure the proper equilibration of the system and also for comparing the efficiency of different Monte Carlo algorithms, we here go beyond temperature annealing schemes, exploring a number of different Monte Carlo methods.  We also use these methods to identify the most challenging factors involved in the thermal equilibration.  The algorithms we have implemented are widely used in spin glasses \cite{EA,SK,young:98}, where systems are highly disordered and frustrated, and equilibration is usually essential for progress.  They include simulated annealing \cite{SA} for optimizations, and parallel tempering \cite{ptmc1,ptmc2,Hukushima:PT} for equilibrium sampling.  One more recent but less known algorithm is the population annealing \cite{Hukushima:PA,Zhou:PA,Machta:PA,Wang:PA,Wang:TBC,Weigel:PA,Amey:PA,Amin:PA}.  Both population annealing and parallel tempering are extended ensemble methods with similar efficiency for spin glasses. We find, however, that parallel tempering can be more efficient than population annealing for our less frustrated model. Using our equilibrium sampling to generate energy histograms for a series of system sizes, we find a single first-order phase transition, from cluster liquid to cluster crystal, featuring a double-peak structure in the histograms.   In addition we studied the case of particles moving in a harmonic trap. 
Also here we obtain a cluster crystal phase, notably with a lattice spacing  
approximately independent of the distance to the trap center and similar to the unconfined case.  However, the occupation number of the clusters decreases with the distance to the center. Accordingly, the melting temperature becomes position dependent with more stable highly populated clusters located at the center of the trap, generating an interesting inhomogeneous melting.

The paper is structured as follows.  First we introduce our model, observables, simulation methods, and the mean-field analysis in Sec.~\ref{mm}.  The algorithms and numerical results are described in Sec.~\ref{results}.  Finally, a summary of the main findings and discussion of their implications is presented in Sec.~\ref{cc}.

\section{Model, observables and methods}
\label{mm}

\subsection{Model and observables}
The two-dimensional system of 
monodisperse particles interacting via an ultrasoft potential is described by the Hamiltonian
\begin{eqnarray}
H = \sum_{i<j} \dfrac{U_0}{1+(r_{ij}/r_c)^6}.
\end{eqnarray}
Here the indices $i, j$ run from 1 to the number of particles $N$. The particles are placed on an $L \times L$ square with periodic boundary conditions, and the density of particles is $n=N/L^2$.
In Sec.~\ref{BEC} we study effect of confinement by adding an external potential $V(r) = \Omega^2 r^2/2$,
and in this case periodic boundary conditions are not applied.
We take $U_0=r_c=1$ without loss of generality, thus measuring temperature in units of $U_0$, length in units of $r_c$, and density in units of $1/r_c^2$. When computing the distance $r_{ij}=(x_{ij}^2+y_{ij}^2)^{1/2}$ and the potential energy, we define $x_{ij}=\min[|x_j-x_i|,L-|x_j-x_i|]$ and similarly for the $y$-direction.
The phase diagram of this model has been presented in Refs.~\cite{Rogelio:Cluster,Rogelio:Glass} in the range $0.8 \lesssim n \lesssim 2.2$, 
along with the mean number of particles per site.
This model can describe, e.g., clustering of vortices in thin films of Type-1.5 superconductors,
or layered structures sharing 
important features with the intervortex forces there \cite{Rogelio:Glass}.
The soft core potential is also relevant for cold atom physics.
For Rydberg states in $^{87}$Rb the interaction
parameters correspond to about 6 nK and 1 $\mu$m~\cite{cinti14}.

Our main observables are the orientational order parameter $\phi_6$ quantifying the ordering of the triangular cluster-crystal phase, and the specific heat $C_V = \beta^2 \times$var$(E)$, where $E$ is the total energy of the system.
The calculation of $\phi_6$ for a given configuration is based the cluster positions and we now discuss how to compute it in detail. 
We first use a hierarchical clustering technique \cite{DataMining,Rogelio:Cluster} that group particles into clusters deterministically. 
Each particle starts as a single cluster, and the two closest clusters are joined together if their center of mass distance is smaller than a chosen cutoff.
We used a cutoff of $0.7 \approx a/2$ but the outcome is not sensitive to this value and $0.6$ or $0.8$ works equally well. 
When two clusters are joined, the new center of mass is updated. 
The process repeats until no further grouping is possible. Note that the process works for both the cluster crystal phase and the liquid phase. In summary, the process takes a particle configuration $\{\vec{r}_i, i=1,2,...,N\}$ and outputs the centers of mass of $C$ clusters $\{\vec{R}_i, i=1,2,...,C\}$. The clustering process is followed by a further Voronoi decomposition to identify neighboring clusters.
The $\phi_6$ is finally defined as
\begin{eqnarray}
\phi_6 = \left| \dfrac{1}{C} \sum_{j=1}^{C} \left( \dfrac{1}{N_j} \sum_{\ell=1}^{N_j} e^{i6\theta_{j\ell}} \right) \right|,
\end{eqnarray}
where $\theta_{j\ell}$ is the angle defined from an arbitrary direction (often the $\hat{x}$-axis) and the vector $\vec{R}_{\ell} - \vec{R}_j$, and $N_j$ is the number of neighbors of the $j_{th}$ cluster. Note that only neighboring pairs are summed. If the system is a perfect triangular lattice then $\phi_6=1$, and if the system has no orientational long-range order order then $\phi_6=0$. Note that this quantity characterizes cluster ordering, not particle ordering.
We also calculate the 2D radial correlation function defined as
\begin{eqnarray}
g(r) &=& \frac{1}{N} \dfrac{\delta n(r)}{2\pi r \delta r},
\end{eqnarray}
where $\delta n(r)$ is the number of particles in the shell $2\pi r \delta r$, with reference to an arbitrary particle.  Note that the function is normalized as $\int_0^{\infty} g(r) 2\pi r dr =1$.  The function shows a peak at the lattice constant when applied to a cluster crystal configuration, and we use this peak to extract the lattice constant of the crystals.

\subsection{The Monte Carlo methods}
We have studied the system using three different Monte Carlo methods: simulated annealing (SA), population annealing (PA), and parallel tempering (PT).  The major and common component for each algorithm is a Monte Carlo sweep. We use the Metropolis algorithm in this work. Each Monte Carlo sweep is a sequential update of all the $N$ particles.  For each update, we propose to shift a particle randomly within a square box of length $2\sqrt{1/n}$ centered on the particle.  We assume readers are familiar with the single-temperature Markov chain, and the SA for sequential cooling through a series of single-temperature Monte Carlo following an annealing schedule.  We additionally use PT for multiple Markov chains with exchange of replicas between different temperatures.  The PA method is, however, relatively new, and we discuss this algorithm in some detail here.

Like PT, 
PA is an extended-ensemble algorithm for thermal equilibrium sampling.  It is also similar to SA but with a large population of $R$ replicas.  It is often the case, also in this work, that replicas are initialized randomly at $\beta=0$ and cooled to a target temperature $T_{\rm{min}} = 1/\beta_{\rm{max}}$.  The replicas traverse an annealing schedule with $N_T$ temperatures by slowly lowering the temperature.  The population is, however, resampled at each temperature step to maintain thermal equilibrium, with a self-consistent reweighting process.  When the temperature is lowered from $\beta$ to $\beta^\prime$, the population is resampled.  The mean number of copies of replica $i$ with energy $E_i$ is proportional to the appropriate reweighting factor, $\exp[-(\beta^\prime-\beta) E_i]$.  The constant of proportionality is chosen such that the expectation value of the population size at the new temperature is $R$.  In practice, the number of each replica is rounded to the floor or ceiling of the expectation value with the right probability, called nearest integer resampling.  The resampling is followed by $N_S$ sweeps to each replica of the new population, and the cycle of resampling and sweeps repeats until the target temperature is reached.

Note that PA and SA have a similar structure. Turning off resampling, PA becomes SA of $R$ independent runs. PA has been found to be similar in efficiency to PT for spin glasses, but has a number of additional useful features such as having intrinsic equilibration measures (see below), giving direct access to free energy using the free energy perturbation method, and is massively parallel.

The equilibration measure of PA in this work is based on the family entropy $S_f$ and the entropic family size $\rho_s$ \cite{Wang:PA,Wang:TBC}.  These quantities are, respectively, defined as
\begin{eqnarray}
S_f &=& -\sum_i \nu_i \ln \nu_i, \\
\rho_s &=& \lim_{R \rightarrow \infty} R/e^{S_f},
\end{eqnarray}
where $\nu_i$ is the fraction of the population that has descended from replica $i$ in the initial population.  Intuitively, $\exp(S_f)$ characterizes the number of effective surviving families or the diversity of the population, and $\rho_s$ characterizes the average effective surviving family size. 
A large $S_f$ ensures accurate sampling of the equilibrium distribution.
On average, $S_f$ decreases with $\beta$ with a rate that depends on the free energy landscape
and the simulation parameters.
We use $S_f \geq \ln(100)$ in this work, which has been used in many systems \cite{Wang:PA,Wang:TBC}.
If this condition is not fulfilled at the final temperature, the simulation is restarted with a larger initial population.
In addition to providing an equilibrium measure, $S_f$ is also useful for identifying bottlenecks of the equilibration.

\subsection{Mean-field analysis of the cluster crystal ground state}
\label{mft}
In this section, we develop a mean-field analysis to capture quantitative features of the cluster crystal phase. For simplicity, we focus on the ground state which has no thermal fluctuations.  In particular, we are interested in how the mean cluster size $c_s$ and lattice constant $a$ change as a function of the particle density $n$. It is interesting to see whether the mean-field treatment can catch the onset of particle clustering. Finally, we also compare the triangular lattice and the square lattice. We emphasize that comparing average energy per particle for different lattices has been a well-established practice, see e.g. recent works of \cite{ca:patterns,CCSH}.  Nevertheless, the application to cluster-forming particles provides some interesting insights on the clustering features.

Let the area of the unit cell be $A$. Then $n = c_s/A$ by definition, and $A=\sqrt{3}a^2/2$ for the triangular lattice and $A=a^2$ for the square lattice. Note that the optimum $c_s$ and $a$ are hence related for a given density $n$. We define a single particle interaction energy for a particle sitting at the origin without loss of generality as
\begin{eqnarray} \label{eps}
                                                                                                                                                                                                                                                                                                                                                              \epsilon(c_s) &=& (c_s-1)U(0) + \sum_{i \neq 0} c_s U(r_i),\\
                                                                                                                                                                                                                                                                                                                                                              \label{eps2}
         &=& \sum_i c_s U(r_i) - U(0),
\end{eqnarray}
where the summation is over lattice sites and $r_i$ denotes the distance from the lattice site $i$ to the origin or the lattice site $0$.  The first and second terms of Eq.~(\ref{eps}) are the intra-cluster and inter-cluster interaction energies, respectively. Minimization of $\epsilon$ with respect to $c_s$ gives the optimum cluster size $c_s$ and lattice constant $a$. In summary, for a given density $n$, we aim to find the global minimum of $\epsilon(c_s)$. For each $c_s$, we first compute the lattice constant $a$ and then numerically sum the single particle interaction energy $\epsilon$. We run this procedure for both the triangular lattice and the square lattice.

\begin{figure}[htb] \begin{center} \includegraphics[width=\columnwidth]{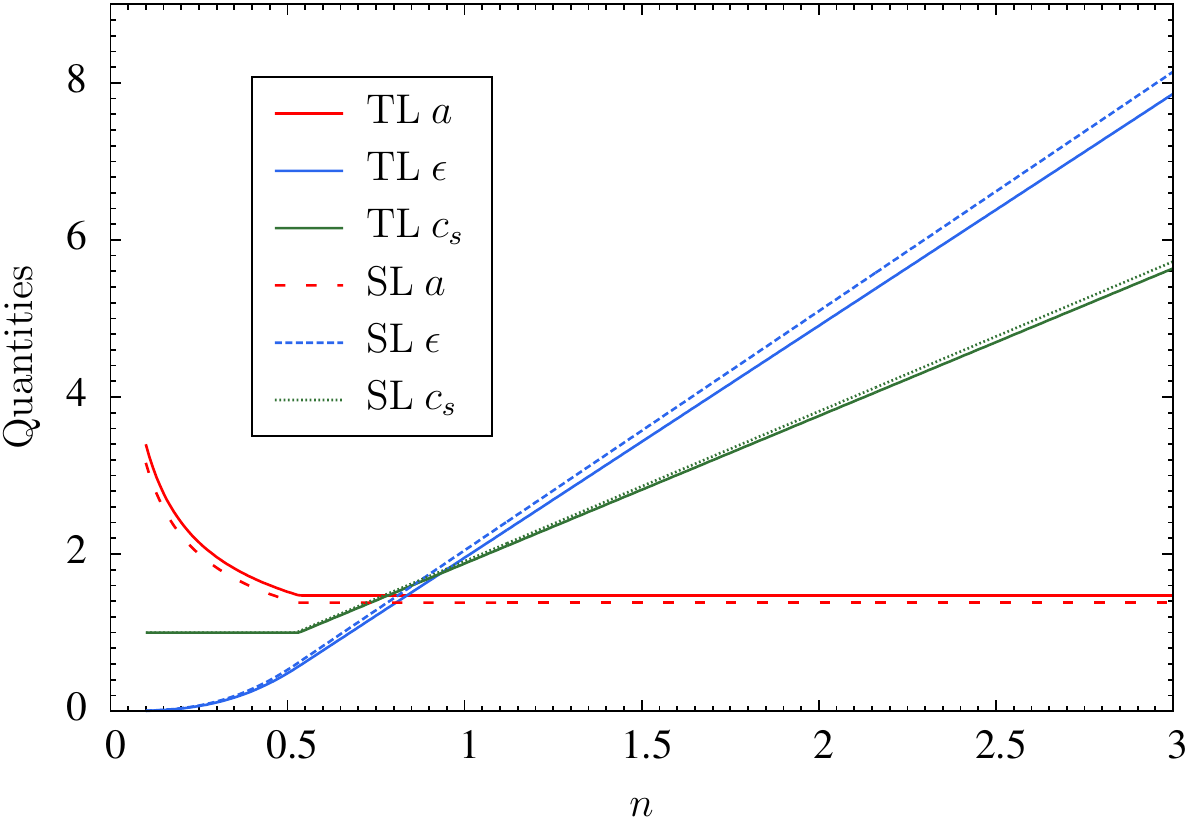} \caption{The cluster crystal lattice constant $a$, single particle interaction energy $\epsilon$, and mean cluster size $c_s$ as a function of density $n$ for both the triangular lattice (TL) and the square lattice (SL) under the mean-field approximation. Note that the lattice constant is essentially independent of the density once the clustering occurs ($c_s>1$) at about $n \approx 0.53$. Therefore, the mean cluster size grows \textit{linearly} with $n$ in the clustering regime. The single particle interaction energy for the triangular lattice is always lower than the square lattice at any finite density, and is also asymptotically linear with density.  }
\label{MFT}
\end{center}
\end{figure}

The optimum single particle interaction energy, lattice constant, and mean cluster size for both lattices are shown in Fig.~\ref{MFT}. 
Firstly, the single particle interaction energy for the triangular lattice is always smaller than the square lattice, and the difference increases with increasing density. Both converge to zero in the zero-density limit.  Second, the mean cluster size has the following onset density of particle clustering at $n_c \approx 0.53$.  When the density is lower than $n_c$, clustering is not favored. When $n \geq n_c$, clustering occurs in the ground state. Once clustering occurs, the lattice spacing remarkably settles down and becomes independent of the density. As a result, the mean cluster size grows \textit{linearly} with density in the clustering regime. This is compatible with the numerical results of Ref.~\cite{Rogelio:Cluster}. The single particle interaction energy is hence also asymptotically linear with density from Eq.~(\ref{eps2}).

\section{Numerical results}
\label{results}
This section has two main parts.  We first present the comparison of different MC algorithms, and the bottlenecks for thermal sampling.  Then we present results for equilibrium properties of the system, including comparison with the mean-field results, the order of the phase transition, and cluster crystals in a parabolic trap.

\subsection{Algorithms}
\begin{figure}[htb] 
\begin{center} 
\includegraphics[width=\columnwidth]{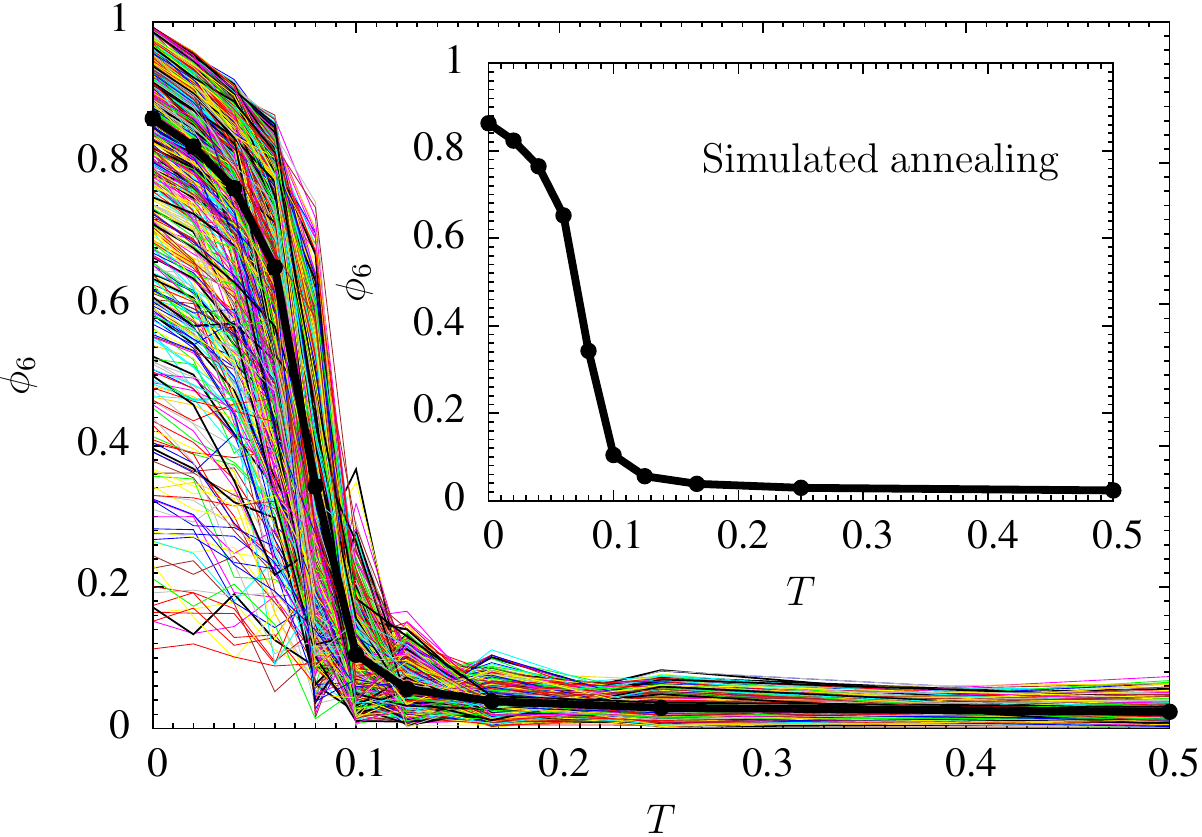} 
\caption{
Evolution of the order parameter $\phi_6$ upon slow cooling using simulated annealing. Each thin line is an independent run and the thick black curve is the mean of these runs which is also highlighted for clarity in the inset panel.
Only 10 temperatures are shown for the mean, although there are 1000 temperature steps.  Note that not all runs go toward $\phi_6=1$, showing that the ground state is a global minimum but not a global attractor. These are 1000 independent runs for $N=1000$ at density $n=1.1$ from $\beta=0$ to $T=0$. We cool the system quite slowly with a total of 1000 steps, with 100 sweeps at each temperature. Half of the temperatures are linear in $\beta$ up to $\beta=10$, and the other half are linear in $T$ to $T=0$. This demonstrates SA can be used for optimization, but is not fully reliable for thermal sampling.}
\label{SA}
\end{center}
\end{figure}

We start by showing that procedures based on simulated annealing cannot easily maintain full thermal equilibration throughout all temperatures even for the clean system.  It might be tempting to expect otherwise, especially considering that after a deep quench from a random configuration, the system can reasonably restore the order parameter close to the transition temperature~\cite{Rogelio:Glass,pc}.  When the system is cooled slowly, one might find the crystal lattice easily, while further lowering the temperature would only suppress lattice fluctuations. However, running a simulated annealing, the system does not always find a perfect lattice, but may contain defects such as grain boundaries near the transition temperature $T_C$. Here, $T_C$ can be estimated as the temperature where $\phi_6$ departs from zero or from the peak of the specific heat. Once they have formed, it may be difficult to remove such defects using local updates. See Fig.~\ref{SA} for an example of a reasonably careful annealing of 1000 independent runs for $N=1000$ particles at density $n=1.1$.  While some fraction of the runs find pretty good lattice structures, the majority of them do not, and some fail badly with little long-range order. The mean, shown by the black curve, is systematically below the top-most curve, which runs toward one when $T=0$. Simple Monte Carlo running at a single temperature, as in quenching dynamics for glass formations, would be only worse.

\begin{figure}[htb]
\begin{center}
\includegraphics[width=\columnwidth]{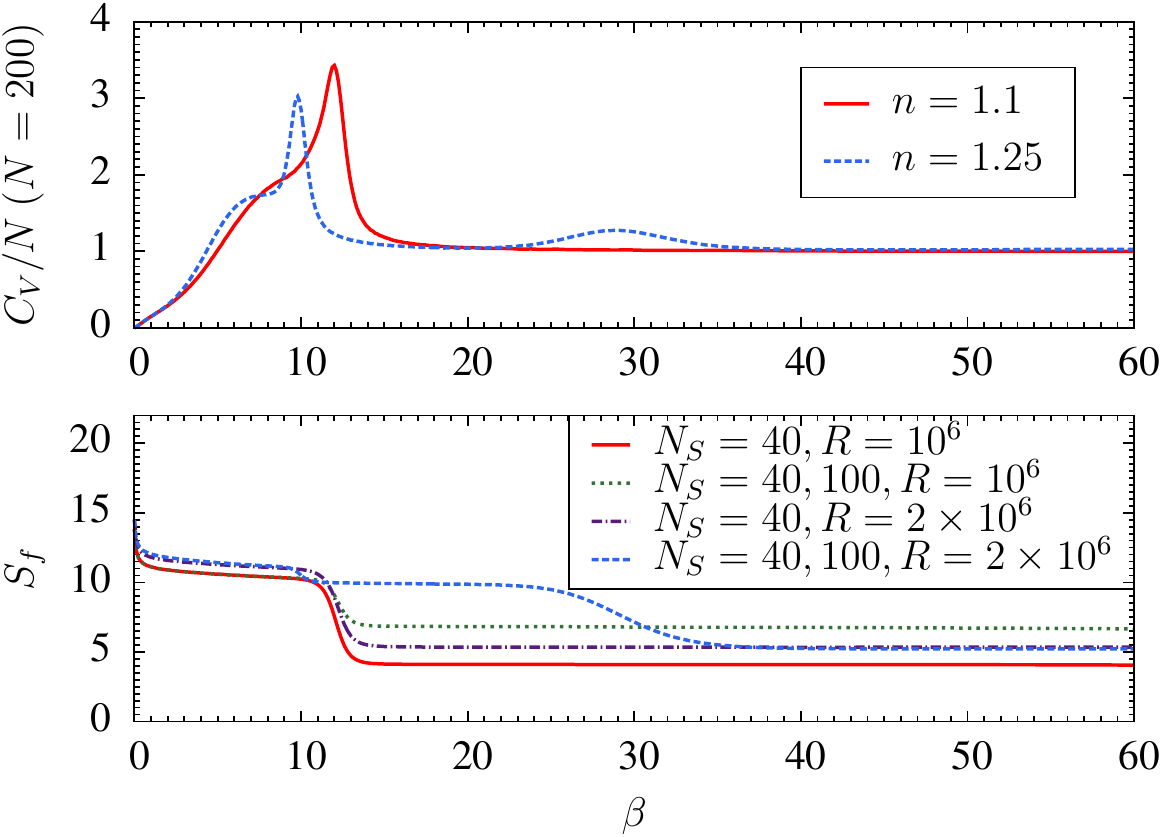}
\put (-214,164) {(a)}
\put (-214,79) {(b)}
\caption{
Specific heat [(a)] and family entropy [(b)] as a function of $\beta$ in two representative scenarios, with and without the low-temperature bump. In both cases, the first-order phase transition is a generic bottleneck, which requires more work near the transition for an efficient sampling. For example, spending 100 sweeps near the transition (40 sweeps at other temperatures) is even more efficient than doubling the system size keeping 40 sweeps at all temperatures. See text for more details. However, there could be an additional bottleneck associated with the anomalous peak at low temperatures, such as for $n=1.25$. Careful inspection shows that this is due to lattice reorganization: there are 86 clusters at higher temperatures below the bump and 90 above. Such bumps at very low temperatures pose even greater equilibration challenges, as shown by the sharp decrease of the family entropy. Furthermore, spending more sweeps here does not help much, in contrast to the phase transition. The shoulder peaks at small $\beta$ are not generic but occur only for small sizes. We use a small size here as we traverse a wide temperature range to $\beta=60$ which is hard for large sizes.}
\label{Cv}
\end{center}
\end{figure}

Nevertheless, comparing the onset of the ordered phase with the peak of the specific heat in Fig.~\ref{Cv}, it seems reasonable that SA captures the transition temperature $T_C$ reasonably well. 
Furthermore, SA remains a valuable tool for optimization to reach very low temperatures (including $T=0$) as thermal equilibration down to very low temperatures could be rather difficult even for small sizes, as we will discuss.  
SA has been used previously to study phase diagrams for other cluster crystal forming potentials~\cite{RA1,RA2}.

Population annealing and parallel tempering, unlike simulated annealing, are both designed for thermal sampling. We first use the family entropy of population annealing to identify bottlenecks in the simulations.  As might be intuitively expected, we find  
that the phase transition (which we will demonstrate is first order in a later section) is a \textit{generic} bottleneck. This is reflected as a sharp decrease of family entropy near the transition.  Clearly, increasing the number of sweeps near the transition can improve equilibration significantly.  See Fig.~\ref{Cv} for an example with $n=1.1$ (solid lines), where we have used an annealing schedule with 100 temperatures linear in $\beta$ up to $T_C$ and 100 ones linear in $T$ in the low temperature part. Notice that if we spent 100 sweeps per replica near the phase transition ($\beta \in [8,16]$), the family entropy drops much less compared with 40 at all temperatures. Furthermore, this is even more efficient than simply doubling the population size keeping $N_S=40$, which involves more total work.  Therefore, we conclude that the phase transition requires more work (either more sweeps or equivalently more temperatures) for efficient thermal sampling.

Note also that from the family entropy of $n=1.25$ (dashed lines) in Fig.~\ref{Cv} the phase transition \textit{may} not be the only bottleneck in thermal sampling, which is accompanied by an anomalous peak in the specific heat at low temperatures.  For this harder case, we have used an annealing schedule with 100 temperatures linear in $\beta$ up to $T_C$ and 200 ones linear in $T$ in the low temperature part.  We have again spent 100 sweeps per replica near the phase transition ($\beta \in [8,12]$) compared with 40 at other temperatures such that the transition is not a significant bottleneck. In this example, the anomalous peak is the major bottleneck.  While this is not a generic peak for different sizes and densities, it occasionally shows up, and since temperatures are very low, spending more sweeps around this peak does not help much in contrast to the transition bottleneck.  By careful inspection, we find that the nature of the bump is due to a subtle lattice reorganization: not in the lattice form but in the number of clusters. There are 86 clusters at higher temperatures before the bump but 90 clusters at lower temperatures after the bump, and between the two the population has a mixture of replicas of both kinds. Note that both lattices are the same triangular crystals, and there is no glassy states here as the replicas are kept in thermal equilibrium. This is clearly a balance of energy and entropy of the two finite lattices. While we believe this is probably a finite-size effect, fitting a lattice in a finite space, it certainly suggests that equilibration below and much below $T_C$ for finite systems may involve subtle differences. For example, if one is running SA for a limited number of independent runs, the true ground state might be missed.

We now focus on the efficiency of population annealing and parallel tempering.  Despite the two algorithms are similarly efficient in spin glasses that are highly disordered, it appears that population annealing is not as efficient as parallel tempering for the clean system. We have also found similar results in our recent simulations of the three-dimensional Ising model. Suppose one is interested in the phase transition regime or below, we think the following factor is limiting the power of population annealing for clean systems: population annealing is a sequential Monte Carlo and requires to start the population at a high temperature. But resampling is only meaningful when the population size $R$ is sufficiently large, hence population annealing necessarily spends lots of efforts in the high temperature regime before reaching the transition temperature.  No initial equilibration is needed if the simulation starts from $\beta=0$, but it takes a number of steps to get close to $T_C$. One can also choose to start at a high but finite temperature directly, but this would require population annealing to equilibrate initially a large number of replicas. In contrast, parallel tempering can equilibrate fast for clean systems (short thermalization times and correlation times) directly near the transition for a much smaller number of replicas, and therefore good or reasonable statics can be obtained more quickly for a fixed computational effort. In summary, we think the sequential preparation is slowing down PA, compared with PT.  Our results are, however, not in disagreement with the similar efficiency in disordered systems. In that scenario, both the thermalization times and correlation times of PT are much longer, driving the efficiencies of PA and PT together. Nevertheless, PA has been very useful in our studies, e.g., in identifying simulation bottlenecks. It provides both sharper and more quantitative information than the exchange probabilities of parallel tempering.  For equilibrium sampling of monodisperse particles in disordered potentials where equilibration is slow, we expect that the massively parallel PA would again become relevant.

Finally, there is an intrinsic complexity in the model itself: the long-range interactions and the associated $O(N^2)$ complexity for one sweep. While it is an option to introduce a cut-off to the potential, the efficiency necessarily decreases with increasing density and is less relevant when the system size is not exceptionally large. Therefore, this is a significant limiting factor on the sizes accessible to equilibrium simulations. We have managed to fully equilibrate about $1000-2000$ particles around $T_C$ in this work, using state-of-the-art computing resources. In contrast, for clean systems with short-range interactions such as the Ising ferromagnets, one can reach orders of magnitude larger sizes.

In summary, the long-range interactions and the phase transition are generic bottlenecks, and extra bumps at low temperatures when present are often a significant bottleneck if one is interested in the low-temperature regime. Increasing the number of Monte Carlo sweeps near the phase transition usually improves efficiency. We find that SA is not sufficient for maintaining full equilibration throughout the annealing, and parallel tempering is more efficient than population annealing for clean systems, although for disordered and frustrated systems the efficiency becomes similar.

\subsection{Equilibrium properties}
We first check the mean-field results and present the phase diagram, followed by a demonstration that the liquid to cluster crystal phase transition is first order, using energy histograms.  Our results for thermal sampling are studied mainly using PT, but some PA results are also presented when relevant.  We find ground states using simulated annealing in the same way as discussed in the previous section, down to $T=0$ with at least 1000 independent runs. The simulation parameters are summarized in Table \ref{table}.

\begin{table}
\caption{
Simulation parameters of some representative examples of the Monte Carlo methods. Here $n$ is the density of particles, $N$ is the number of particles, $R$ is the number of replicas, $N_T$ is the number of temperatures, $N_S$ is the number of sweeps for each replica at each temperature, $\beta_{\rm{min}}$ and $\beta_{\rm{max}}$ are the minimum and maximum inverse temperatures, respectively, and $M$ is the number of independent runs.}
\label{table}
\begin{tabular*}{\columnwidth}{@{\extracolsep{\fill}} l c c c c c c c r}
	\hline
	\hline
	Method &$n$ &$N$ &$R$ &$N_T$ &$N_S$ &$\beta_{\rm{min}}$ &$\beta_{\rm{max}}$ &$M$  \\
	\hline
	SA &1.1 &1000 &$100$ &1001 &100 &0 &$\infty$ &10 \\
	PA &1.1 &200 &$10^6$ &201 &40 &0 &60 &6 \\
	PA &1.25 &200 &$2 \times 10^6$ &301 &40 &0 &60 &2 \\
	PA &1.5 &200 &$10^6$ &201 &60 &0 &60 &6 \\
	PA &1.75 &200 &$10^6$ &201 &40 &0 &60 &2 \\
	PA &2 &200 &$10^6$ &201 &60 &0 &60 &6 \\
	PA &2.5 &200 &$10^6$ &201 &40 &0 &60 &2 \\
	PA &3 &200 &$10^6$ &201 &40 &0 &40 &2 \\
	PT &1.1 &200 &$-$ &64 &$2.2\times10^6$ &5 &20 &2 \\
	PT &1.1 &400 &$-$ &64 &$3.3\times10^6$ &10.5 &11.5 &2 \\
	PT &1.1 &800 &$-$ &64 &$3.3\times10^6$ &10.5 &11.5 &2 \\
	PT &1.1 &1200 &$-$ &64 &$3.3\times10^6$ &11 &11.5 &2 \\
	PT &1 &1000 &$-$ &64 &$3.3\times10^6$ &10 &16 &2 \\
	PT &1.5 &1000 &$-$ &64 &$3.3\times10^6$ &4 &8 &2 \\
	PT &2 &1000 &$-$ &64 &$3.3\times10^6$ &2 &6 &2 \\
	PT &2.5 &1000 &$-$ &64 &$3.3\times10^6$ &1 &5 &2 \\
	PT &3 &1000 &$-$ &64 &$3.3\times10^6$ &0.5 &4 &2 \\
	\hline
	\hline
\end{tabular*}
\end{table}

The $T_C$ is estimated from the peak of the specific heat.  We extract from the ground state the lattice constant and the mean cluster size to compare with the predictions of the mean-field theory.  Ideally we should find well ordered ground states with $\phi_6 \geq 0.99$. While this is readily achievable for large densities, it turns out to be quite a difficult task for low densities where the transition temperatures are very low and particles interact only very weakly. We therefore require this criterion only for densities of $n \geq 1$.  Nevertheless, we find that the states still have reasonably good local order, despite of the poor global order.  Remarkably we find that the location of the first nontrivial peak of the correlation function does not significantly depend on whether the system is in the true ground state or just a low-energy state, and often even a glassy state.

\begin{figure}[htb] 
\begin{center} 
\includegraphics[width=\columnwidth]{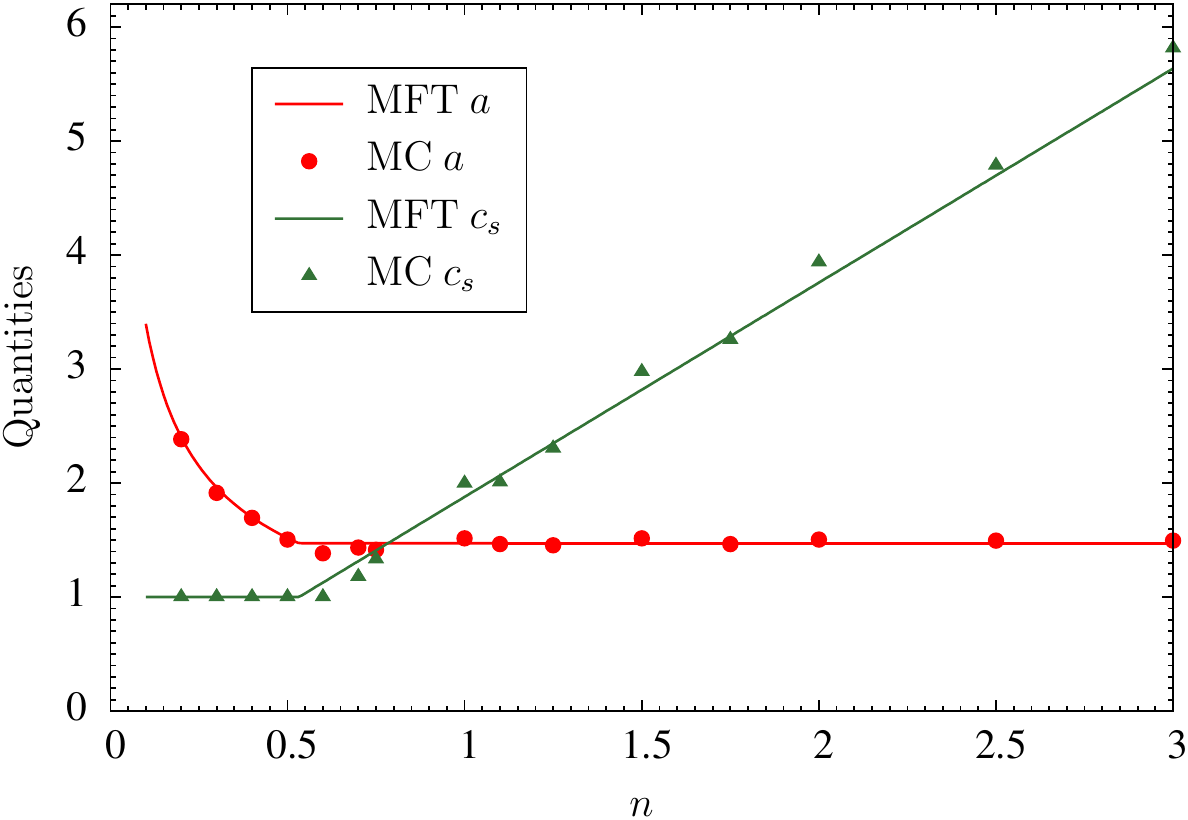}
\put (-16,28) {(a)} \\
\includegraphics[width=\columnwidth]{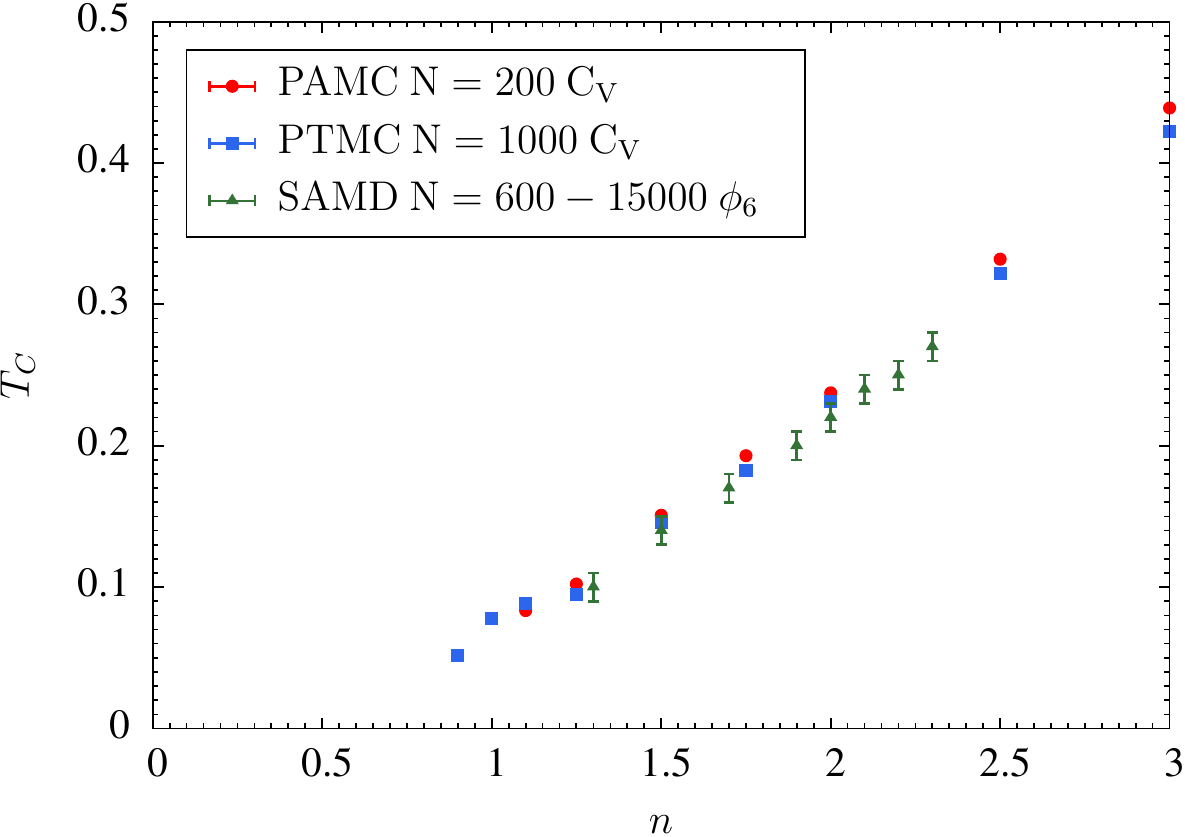}
\put (-16,28) {(b)}
\caption{ Top panel: Comparison of the mean-field and simulated annealing results characterizing the ground states [(a)]. Bottom panel: Phase diagram estimated from the specific heat peaks of the equilibrium MC simulations, and the estimation of simulated annealing using molecular dynamics and $\phi_6$ \cite{Rogelio:Glass} [(b)]. Errorbars of MC results are smaller than the symbols. 
		}
\label{MFT2}
\end{center}
\end{figure}

Our results are summarized in the top panel of Fig.~\ref{MFT2}. We find the mean-field analysis is in reasonably good agreement with the numeric results. The system has an onset density, and the lattice constant is independent of density in the cluster-crystal region.  Below the onset density, the MFT is trivially exact as $c_s=1$. The MFT, however, does not find the critical onset density exactly.  The numerical onset density appears to be between $0.6$ and $0.7$, while the MFT predicts $0.53$.  The asymptotic lattice constant $a=1.495$ found in the simulations at the two largest densities shows a very good agreement with the MFT result $a=1.473$.  The small discrepancy is due to neglecting of correlations in our MFT approximation.  For example, clusters of different sizes are likely not randomly distributed \cite{Rogelio:Glass}.  There is also place for finite-size effects on the Monte Carlo side, as we cannot exclude that for large sizes the MC results will get closer or even converge to the MFT prediction.  Nevertheless, the results are reasonably close, and from our MFT analysis, we can clearly see that the onset of particle clustering is a process of minimization of the lattice energy at low temperatures.

The phase diagram is summarized in the bottom panel of Fig.~\ref{MFT2}. We have identified $T_C$ as the location of the maximum peak of the specific heat. This again works well for high densities $n \gtrsim 1$. There are small finite-size effects as the peaks of $N=200$ and 1000 are not at the same temperature, but they are sufficiently close to get an estimate of the phase boundaries. Our results regarding the melting temperature are in good agreement with those obtained from the evolution of $\phi_6$ using simulated annealing~\cite{Rogelio:Glass,Rogelio:Cluster}.

A region of phase coexistence can be present at a first order melting of an ordinary particle crystal, where the density jumps discontinuously at the transition when simulated in an NVT ensemble. 
For a cluster crystal, on the other hand, the lattice spacing can readjust without changing the total density by changing the cluster occupation numbers, and for this reason it is not clear that a phase coexistence region necessarily exists.
In fact, our simulation results do not clearly indicate a coexistence regime in the phase diagram,
suggesting that such a region might be very narrow or absent in the model investigated here.

\begin{figure}[htb]
\begin{center}
\includegraphics[width=\columnwidth]{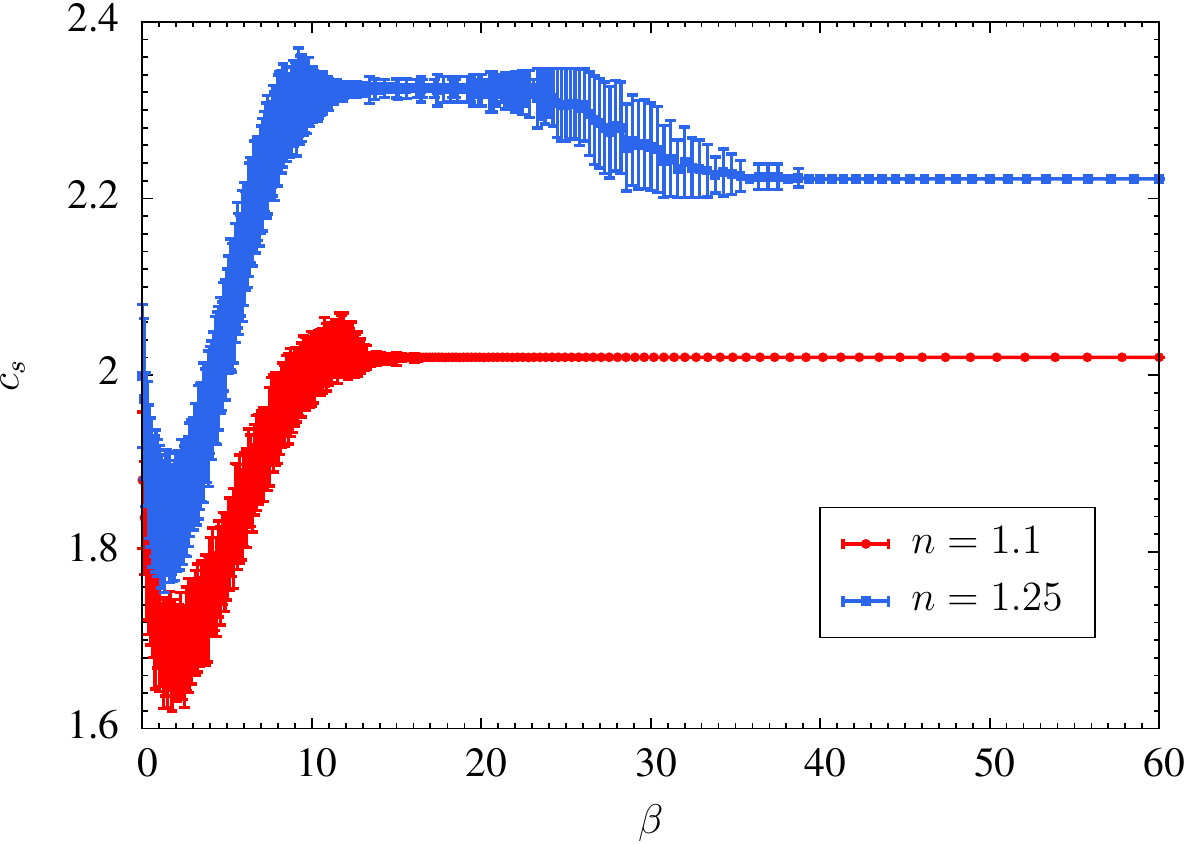}
\caption{
The corresponding mean cluster size $c_s$ as a function of temperature for Fig.~\ref{Cv}. The density $n=1.25$ shows clearly a change of $c_s$ at the anomalous peak. Other than this finite size effect, the cluster size does not depend on the temperature in the crystal phase for this ultrasoft model. The errorbar here is for the spread of the distribution with each point estimated from $100$ configurations, not the errorbar of the mean of $c_s$ which is smaller by a factor of $\sqrt{100}=10$. Note, however, the clustering procedure or counting of clusters is most reliable only at low temperatures. See the text for more details.
}
\label{cs}
\end{center}
\end{figure}

It is interesting to know to what extent the lattice parameters apply to finite temperatures. It is clear that the identity and hence counting of clusters is less precise at finite temperatures, due to particle hopping among clusters. However, we can still look for a trend and study if there is an obvious change in the size of clusters, following our clustering method. Such results are particularly relevant in the low-temperature limit. Therefore, we go back to the examples in Fig.~\ref{Cv} and the temperature evolution of the mean cluster size $c_s$ is shown in Fig.~\ref{cs}. We see clearly the fluctuations of the mean cluster size from configuration to configuration is essentially zero in the $T \rightarrow 0$ limit, and other than the finite-size effect at the anomalous peak, the mean cluster size approximately does not depend on the temperature in the low temperature limit. This even holds close to the transition temperature, though one should be cautious on the definition and counting of clusters in this regime. Nevertheless, this clearly shows the ground state lattice constant remains a relevant length scale throughout the solid phase. We show that this is actually true even beyond the phase transition in the liquid phase, 
using pair correlation functions in Sec.~\ref{cl}.

The order of this phase transition between the fluid and the cluster-crystal region has not been conclusively investigated previously.  It has been reported that this transition occurs in a single step for the generalized exponential model of index 4 (GEM4), where the transition was encountered to be of first-order nature~\cite{prestipino14}. Using density functional theories and simulations a transition in the ordering of clusters but not for particles was found. Here we use equilibrium sampling to generate energy histograms of our model (similar to GEM4) to directly demonstrate the discontinuous nature of the transition. Our conclusion is therefore in agreement with Ref.~\cite{prestipino14}.
 
\begin{figure}[htb] \begin{center} 
\includegraphics[width=\columnwidth]{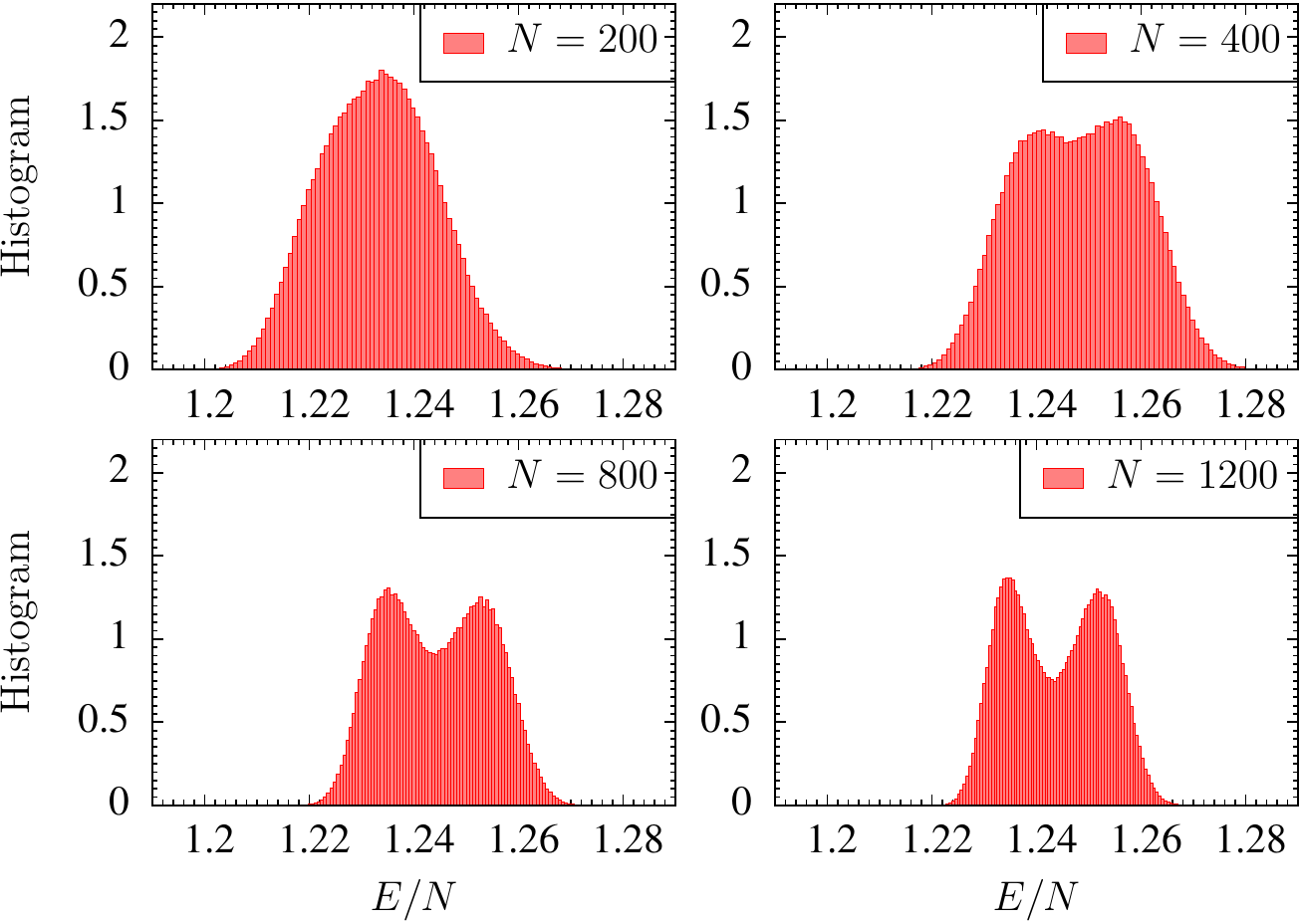}
\put (-214,28) {(c)} 
\put (-214,110) {(a)} 
\put (-96,110) {(b)} 
\put (-96,28) {(d)}\\
\includegraphics[width=\columnwidth]{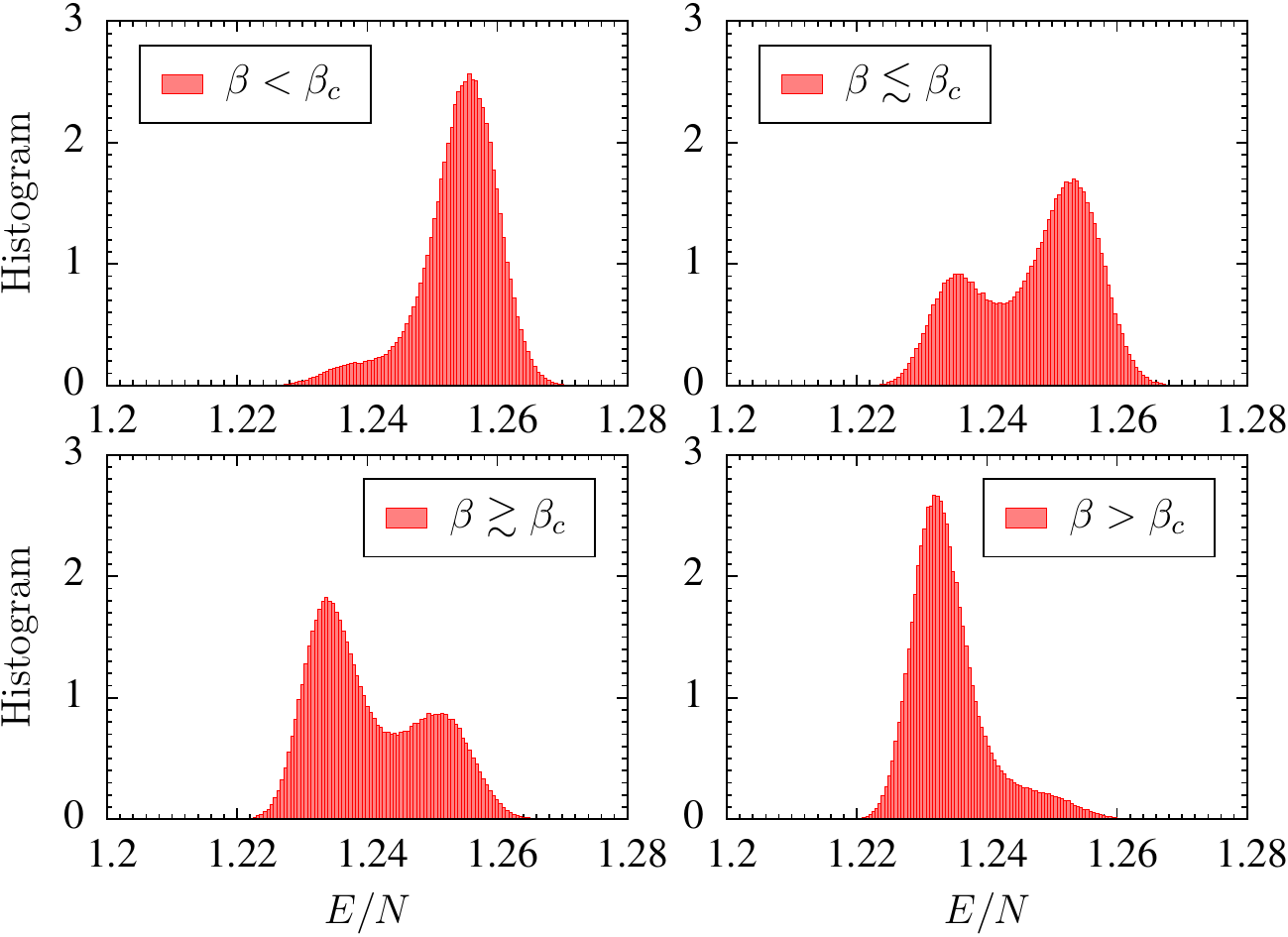}
\put (-223,28) {(g)} 
\put (-223,110) {(e)} 
\put (-105,110) {(f)} 
\put (-105,28) {(h)}
\caption{ Top panel: Energy histogram at the phase transition for a series of sizes $N=200, 400, 800$ and $1200$, respectively [(a-d)] at density $n=1.1$. There is a double-peak structure emerging with an increasing free energy barrier with increasing size $N$, suggesting that the liquid to cluster-crystal phase transition is first order. Bottom panel: Energy histogram for four selected temperatures before and after the phase transition of $N=1200$. The selected temperatures are $\beta= 11.0806, 11.1774, 11.2419, 11.3347$ [(e-h)], where the critical temperature $\beta_c=11.2097$.}
\label{EH}
\end{center}
\end{figure}

\begin{figure}[htb] \begin{center} \includegraphics[width=\columnwidth]{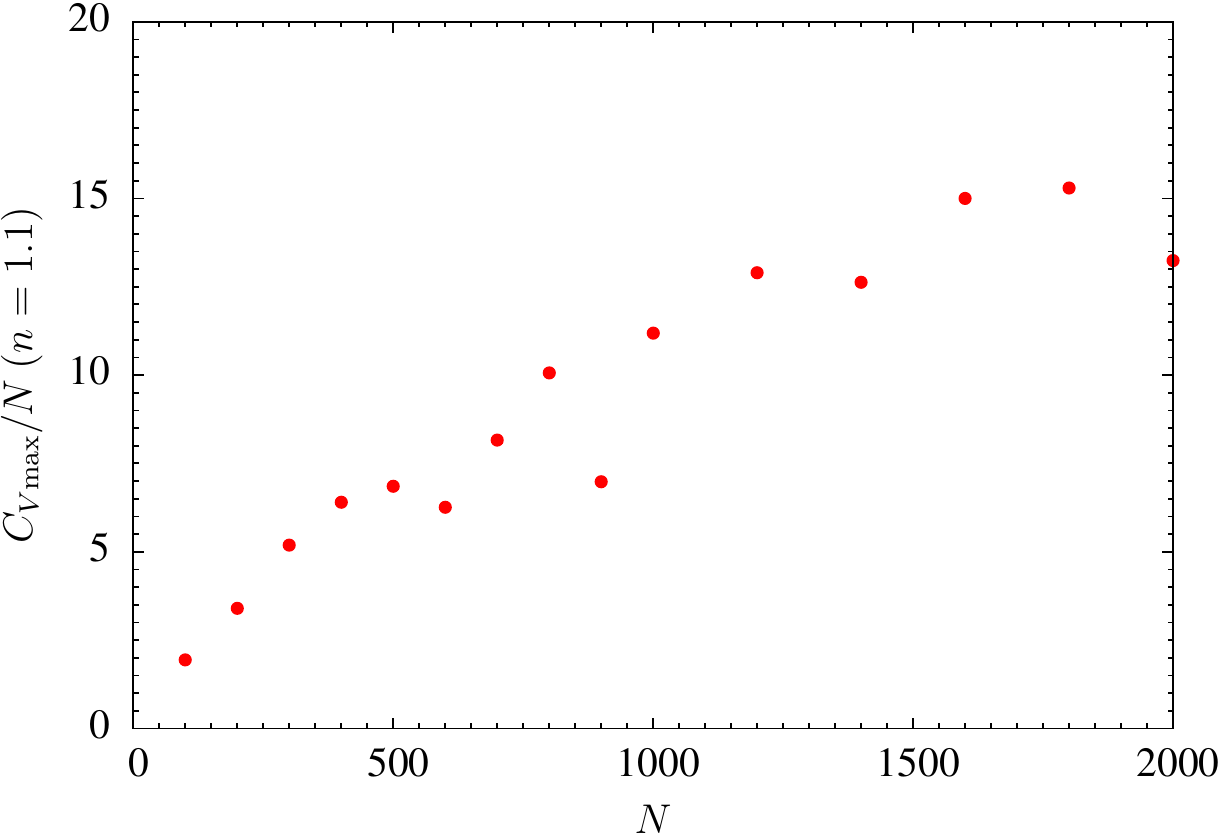} \caption{ Maximum of the specific heat per particle $\max(C_V)/N$ as a function of $N$ at the density $n=1.1$. We see after an initial transient at small sizes, the function develops approximately into a linear function, compatible with a first-order phase transition. Errorbars are smaller than the symbols.  }
\label{Cv2}
\end{center}
\end{figure}

The energy histograms for several sizes near $T_C$ at density $n=1.1$ are shown in Fig.~\ref{EH}.  We have refined our temperature set to find symmetric distributions, which require a very fine temperature spacing.  While there is still a bit of asymmetry, a clear trend can be observed in the energy histograms, where a double-peak structure appears with a growing free energy barrier for increasing system size.  This is a strong signature of a first-order phase transition.  In addition, we have measured the maximum value of $C_V/N$, shown in Fig.~\ref{Cv2}.  We see that after an initial transient at small sizes, the function develops an approximately linear behavior, compatible with a first-order phase transition.  Again, $C_V/N$ does not include any cluster parameter, so we conclude the liquid to cluster-crystal phase transition is first order, in contrast to the KTHNY theory. While there is no theoretical basis to exclude the possibility of two transitions (either two second order or two first order, or one first order and one second order), we do not see any evidence that this occurs in our simulations of the ultrasoft potential.

\subsection{Cluster liquids}
\label{cl}
\begin{figure}[htb] \begin{center}
\includegraphics[width=\columnwidth]{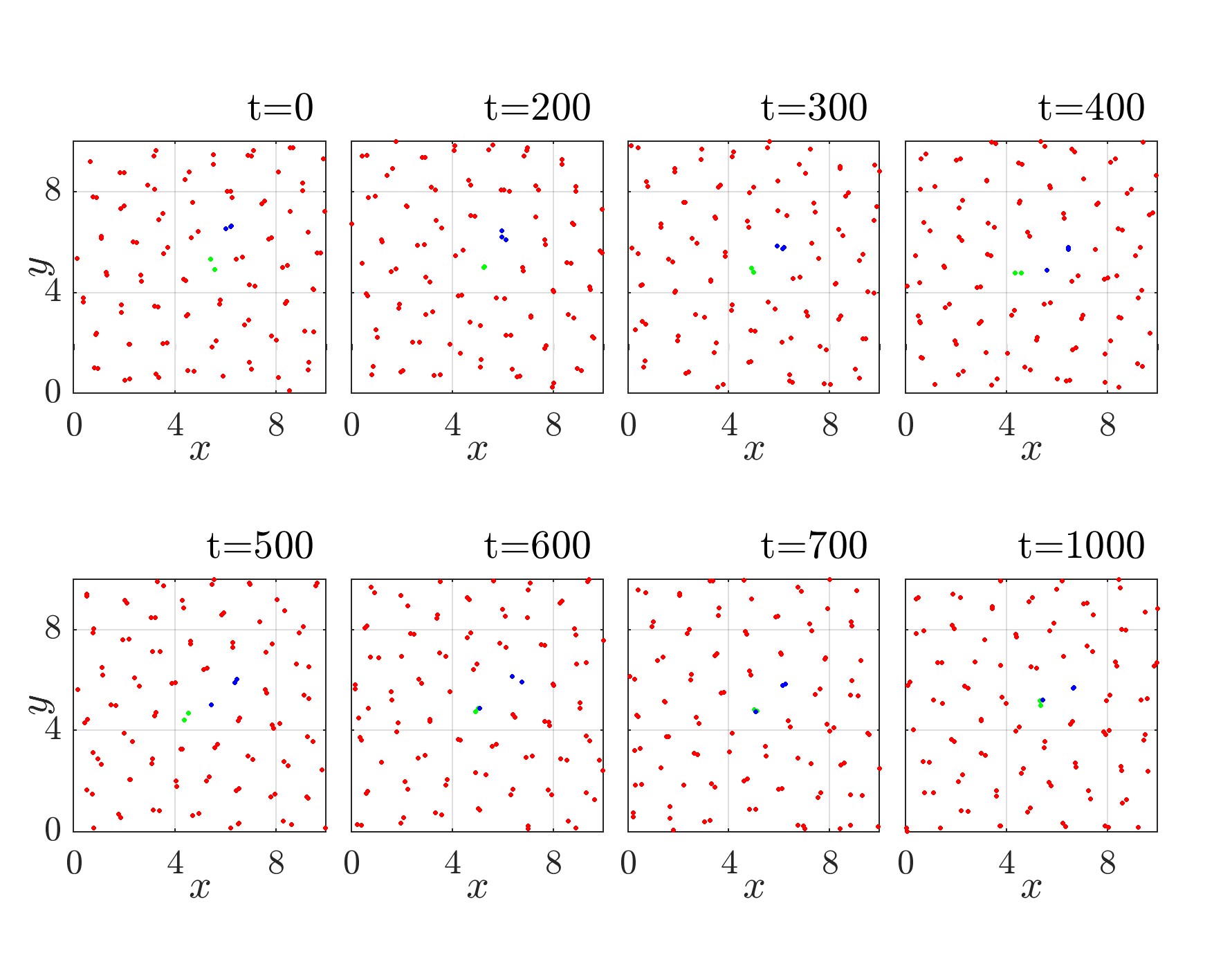} 
\put (-192,100) {(a)}
\put (-136.3333,100) {(b)} 
\put (-80.6667,100) {(c)}
\put (-25,100) {(d)}
\put (-192,12) {(e)}
\put (-136.3333,12) {(f)} 
\put (-80.6667,12) {(g)}
\put (-25,12) {(h)}
\caption{ Typical cluster liquid dynamics for the ultrasoft potential at density $n=1.1$ and $\beta=10$ [(a-h)]. The clusters can oscillate and diffuse for a relatively long time before a particle hops between clusters. In this example, five particles are highlighted. From $t=0$ to $t=300$, they form two clusters. At about $t=400$, the blue cluster loses a particle and forms a single-particle cluster. This cluster later on at about $t=600$ joins the green cluster. The times are in units of Monte Carlo sweeps. See the relevant movie and comparison with a particle liquid in Ref.~\cite{comment:movie}.}
\label{Dynamics}
\end{center}
\end{figure}

The clustering of particles in the cluster crystals remarkably persists to the liquid phase, and hence we here propose the concept of a cluster liquids associated with melted cluster crystals.
It is relatively straightforward to recognize cluster liquids
from some typical configurations shown in Fig.~\ref{Dynamics}. Despite this clearness in our present case, a generic definition turns out to be difficult due to the rich phase diagrams of various potentials. Here, we discuss this only qualitatively and restrict our discussions to simple cases where the associated crystal phases are simple crystals like the triangular lattice, excluding more exotic ones such as stripes. The counterpart of corresponding concepts such as stripe liquids can be realized.

\begin{figure}[htb] \begin{center} \includegraphics[width=\columnwidth]{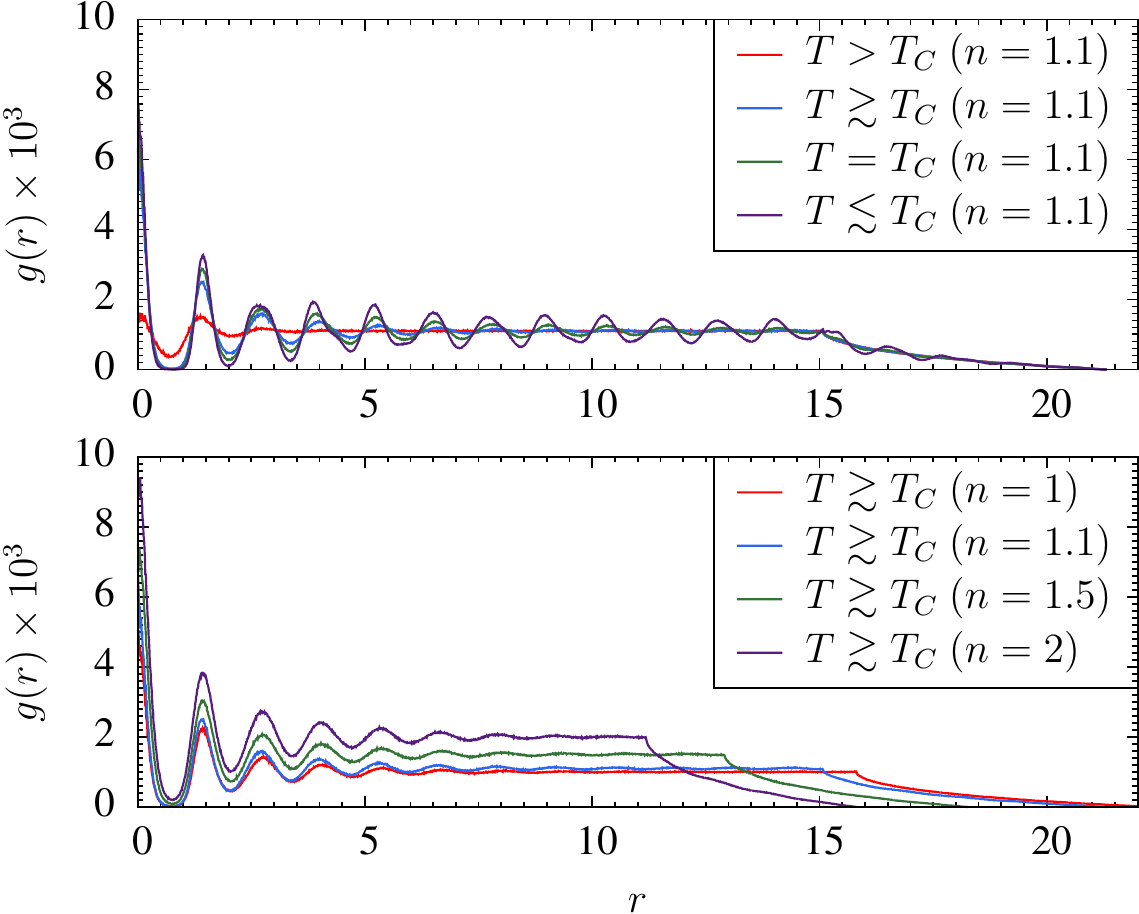}
\put (-212,182) {(a)}
\put (-212,88) {(b)}
\caption{ Top panel: The first few peaks of the pair correlation function occur at the same locations for cluster liquids and cluster crystals [(a)]. Bottom panel: These peaks in cluster liquids are density-independent like the density-independent lattice constant of cluster crystals [(b)]. Note the second peak from left is the cluster lattice constant peak, its height grows when $T$ is lowered [(a)] or when $n$ is increased at fixed $N$ [(b)]. The figure demonstrates that clustering occurs already in the liquid phase and not at the phase transition. In all cases, $\beta = 4, 10, 11.21, 12, 11, 10, 6, 4$, respectively.  }
\label{GR}
\end{center}
\end{figure}

To have a well defined cluster liquid, the liquid phase should preserve some of the cluster order of the solid phase. The configurations in Fig.~\ref{Dynamics} motivate us to look at the pair correlation function. If the particles are indeed clustering in the liquid phase, the first few relevant crystal peaks and in particular the cluster lattice constant peak should appear in the correlation function due to the short-range order of liquids. This is indeed the case, as shown in Fig.~\ref{GR} where the correlation functions for density $n=1.1$ are shown near the phase transition. The second peak from left is the lattice constant peak which is almost temperature independent. This peak persists deep into the liquid phase but becomes less pronounced. This indicates that particles are clustering before solidification to cluster crystals, and clustering does not appear suddenly at the phase transition. Since the lattice constant is independent of density for cluster crystals, we expect that this peak in the liquid phase is also density independent, which is the case as shown in Fig.~\ref{GR}. Therefore, the cluster liquid is well defined.

Having defined cluster liquids, we discuss the major differences of such liquids and particle liquids in terms of their typical dynamical processes. In particle liquids, the dominant processes are particle oscillations in the potential formed by the neighbors and the eventual diffusion of the particles. These two processes also occur in cluster liquids for the centers of mass of clusters. However, due to the clustering, the potential well is much stronger and remarkably particles in a cluster can undergo many oscillations within the cluster diameter before the cluster occupation number changes.
The diffusion of clusters is also much harder than single particles as the diffusion of a cluster is now the center of mass of several particles.
This might be related to our observation of a single first-order phase transition. Note that many such oscillations is necessary for well-defined cluster liquids. This is because particles may also ``cluster'' together by chance for particle liquids. However, such ``clusters'' break almost immediately after forming and as such they are different from the clusters in cluster liquids.
In addition to the two processes above, cluster liquids have also particle exchanges among clusters or the constant dynamical breaking and forming of clusters. There are two typical elementary processes, a cluster may emit or absorb a particle, or two emitted single particles form a new cluster. The combination rate of the latter depends strongly on the density of the system. For low densities such as $n=1.1$, a single particle may be relatively long-lived as a single-particle cluster. Nonetheless, such single particles are the most ``reactive'' in the liquid, having higher mobilities and strong tendencies to form clusters. For higher densities such as $n=2$, an emitted particle joins another cluster much faster, which appears effectively as a particle hopping among clusters.

Next we present typical dynamics of a particle liquid and a cluster liquid. We use here a Gaussian potential (GEM2) and the ultrasoft potential for the two cases, respectively. We have used $N=1000$ particles at the same density $n=1.1$ and $\beta=10$ for comparison, and an updating length scale for a particle $\pm 0.2$ for reasonably time scales of the dynamics. Note that this length scale is still relatively small compared with the lattice constant. In Fig.~\ref{Dynamics}, cluster oscillations along with a particle emission and a subsequent combination are shown for cluster liquids. For more details of both liquids we refer to a movie comparing the two cases \cite{comment:movie}. In both cases, the times are in the unit of sweeps. Finally, it should be emphasized that for MC dynamics to be realistic, our dynamics is mostly relevant to experimental settings where inertia effects are negligible, i.e., the dynamics is heavily overdamped similar to the molecular dynamics simulations of Refs.~\cite{Rogelio:Cluster,Rogelio:Glass}.

\subsection{Confinement effects}
\label{BEC}

Finally, we have also studied the cluster-forming particles in presence of a global parabolic trap, motivated by magnetic trap experiments of the atomic Bose-Einstein condensates \cite{henkel12,BECs}. More specifically, the particles interact with each other and also with an external trap of the form $V(r) = \Omega^2 r^2/2$.  We have explored values from $\Omega=0.1$ to $2$, and obtain triangular cluster crystals for $\Omega \gtrsim 0.4$ at low temperatures. Note that density as well as periodic boundary conditions are no longer relevant in this setting, as particles are confined by the external trap. The trapping frequency $\Omega$ controls the density and size of the system.

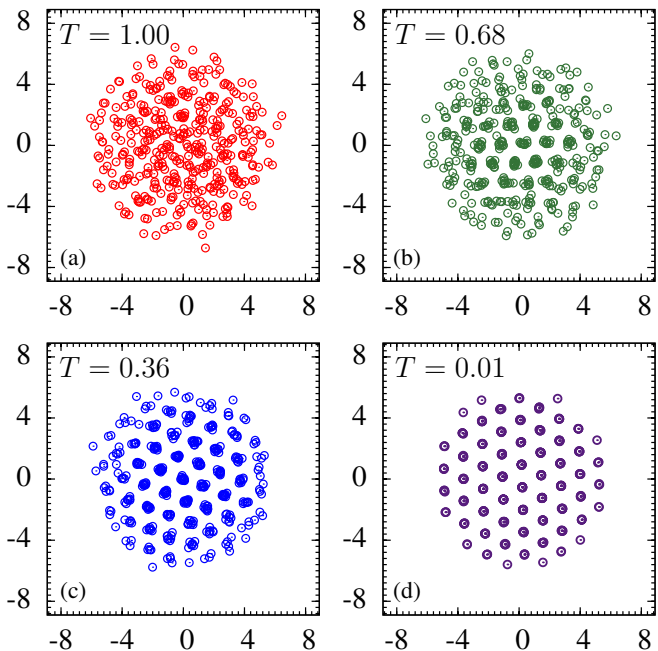
\begin{figure}[htb] \begin{center} \includegraphics[width=\columnwidth]{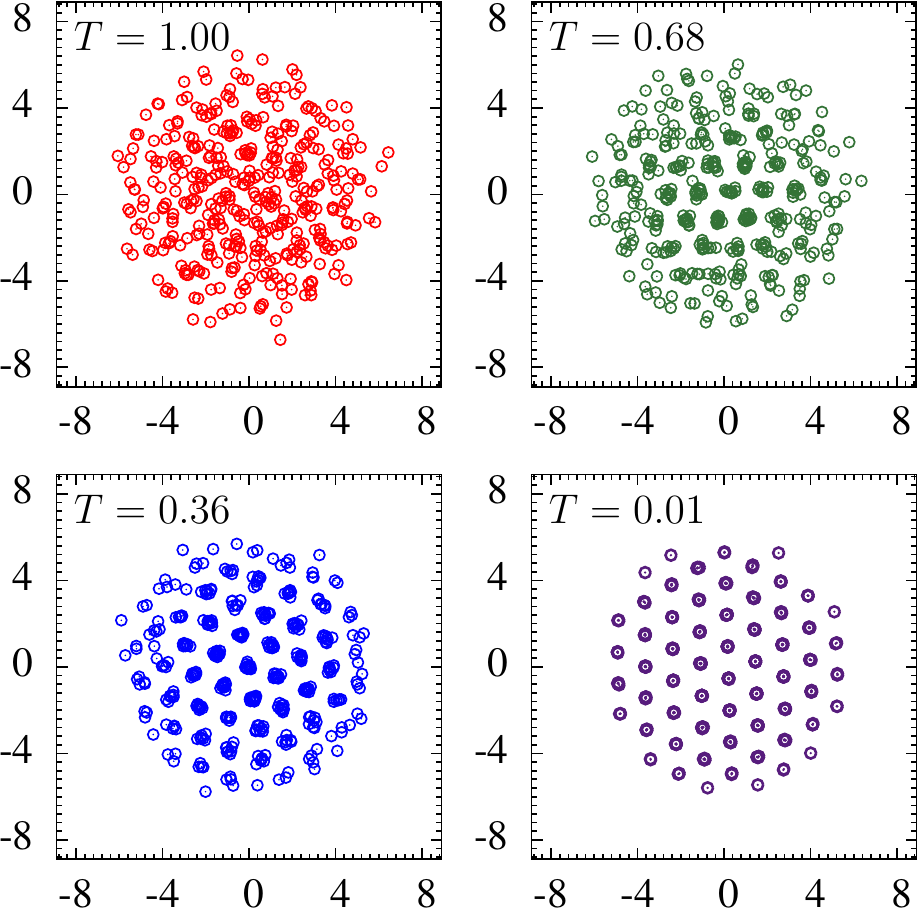}
\put (-226,146) {(a)} 
\put (-100,146) {(b)} 
\put (-226,20) {(c)} 
\put (-100,20) {(d)} 
\caption{ Monodisperse particles in an external harmonic trap of strength $\Omega=1$ with 400 particles [(a-d)]. Note the same lattice structure as the free case, but with very uneven cluster sizes. The clusters are larger at the center of the trap, and smaller at the edge of the lattice. The broad range of cluster sizes leads to nonuniform crystallization or melting, as larger clusters order first followed by smaller clusters upon further cooling. In particular, at intermediate temperatures such as $T=0.68$, the system remarkably has a solid core with a shell of liquid fluid around it.  }
\label{XYtrap}
\end{center}
\end{figure}

Interestingly, the lattice spacing is again robust in this setting for the range of trapping frequencies studied here.  As $\Omega$ increases, the lattice constant does not change significantly as in non-cluster-forming systems~\cite{puente:tr,ghosal:cc}, but rather particles form larger, more populated clusters, and hence a smaller lattice for the same number of particles.  See Fig.~\ref{XYtrap} for typical equilibrium states at four representative temperatures for $\Omega=1$. The lattice constant of about 1.465 is very similar in all cases.  The cluster sizes are no longer uniform due to the trap. Instead they are larger at the center of the trap and smaller at the edge. The problem is also related to the intervortex potentials with long-range attraction and multiple intermediate repulsive length scales, which is expected in layered systems \cite{meng16}.

As a consequence
the system has many energy scales.  Upon a slow cooling the particles order first at the center of the trap, and then gradually in smaller clusters at the edge; c.f. with a discussion of melting of ordinary vortex lattices in a trap \cite{Egor:TFVM,Egor:TFVM2}. This can be understood from the phase diagram shown in Fig.~\ref{MFT2}.  For intermediate temperatures such as $T=0.68$, the system can remarkably have a solid cluster-crystalline core surrounded by a cloud of fluid.  This inhomogeneous melting is an interesting phenomenon that can be experimentally realized.

\section{Conclusions}
\label{cc}
In this work we studied a system of monodisperse particles interacting via an ultrasoft potential in 2D.
We find that simulated annealing is not effective to reach thermal equilibrium even for this clean system without disorder, while it is still a useful tool for optimizations.  Under these conditions parallel tempering is shown to be more efficient than population annealing.  We have presented a simple mean-field characterization of the ground states and compared its predictions with Monte Carlo results, finding reasonably good qualitative and quantitative agreement.

With the resolutions that we can achieve, the cluster liquid to cluster crystal phase transition is first order and occurs in a single step. We do not find indications for formation of an intermediate hexatic phase, which in contrast is the case for non-cluster-forming systems.  Our analysis is based on equilibrium energy histograms near the transition, without using any cluster-related parameter. The results are relevant for the problem of melting transitions in systems of soft particles and the problem of vortex melting phase transition in superconductors with multi-scale cluster-forming inter-vortex forces \cite{Rogelio:Glass,meng16,babaev2017type}.

\acknowledgments

W.W.\ and E.B.\ acknowledge support from the Swedish Research Council Grant No.~642-2013-7837 and the Goran Gustafsson Foundation for Research in
Natural Sciences and Medicine. 
M.W.\ and R.D.M.\ acknowledge support from the Swedish Research Council Grant No.~621-2012-3984. 
The computations were performed on resources
provided by the Swedish National Infrastructure for Computing (SNIC)
at the National Supercomputer Centre (NSC) and the High Performance Computing Center North (HPC2N).

\bibliography{Refs}

\end{document}